\documentclass[12pt]{article}
\usepackage{amsmath, amssymb}

\begin{document}
\vspace*{-.6in} \thispagestyle{empty}
\begin{flushright}
UK-06-01
\end{flushright}
\baselineskip = 18pt

\vspace{1.0in} {\Large
\begin{center}
{Loop Corrected D-brane Stability Conditions}
\end{center}} \vspace{.5in}

\begin{center}
Chengang Zhou \\
\emph{Department of Physics and Astronomy\\
University of Kentucky, Lexington, KY  40506, USA\\
czhou@pa.uky.edu}
\end{center}
\vspace{0.4in}

\begin{center}
\textbf{Abstract}
\end{center}
\begin{quotation}
\noindent In type-II string theory compactifications on Calabi-Yau
manifolds, topological string theory partition functions give a
class of exact F-terms in the four dimensional effective action. We point out that
in the background of constant self-dual field strength, these terms
deform the central charges for D-branes wrapping Calabi-Yau manifold
to include string loop corrections. We study the corresponding loop
corrected D-brane stability conditions, which for B-type branes at
the large volume limit implies loop corrected Hermitian-Yang-Mills
equation, and for A-type branes imply loop corrected special
Lagrangian submanifold condition.
\end{quotation}

\newpage

\pagenumbering{arabic}


\section{Introduction}

BPS states from D-branes wrapping curved space, in particular
Calabi-Yau manifold, have been among the most important subjects of
string theory research ever since the second string theory
revolution. The effective D=4 $\mathcal{N}=1$ theories
obtained from such brane configuration occupy a large sector of the
string theory phenomenology and provide promising candidates for
reality. The mathematical structures obtained from the brane picture
is quite rich, for example the Kontsevich mirror symmetry conjecture. Also they provide a microscopic
picture for BPS blackholes and many extraordinary results like blackhole entropy counting have been
obtained.

The physics of D-branes wrapping Calabi-Yau has distinctive feature
of the decoupling of the K\"{a}hler moduli and the complex structure
moduli at generic points of the moduli space.  In the effective D=4
$\mathcal{N}=2$ supergravity coupled with matters, one moduli
underlies the vector multiplets and the other underlies the
hypermultiplets, and the decoupling of these two moduli at generic points
of the moduli space is the requirement of the $\mathcal{N}=2$
supersymmetry. From the super-Yang-Mills theory on the branes
wrapping the internal space and expanding the four-dimensional
spacetime, one moduli underlies the superpotential while the other
underlies the supersymmetric D-term, depending on either IIA or IIB
branes. This is the basis of the category description of the all
type II D-branes, as pioneered by M. Douglas and
collaborators~\cite{DFR, D}, that the BPS branes are objects in a
category determined by complex structure for B-type branes and
K\"{a}hler structure for A-type branes. Then the grading is defined
at every point in the other half of the moduli space, i.e. the
K\"{a}hler moduli for B-type branes and complex structure moduli for
A-type branes, and used to define stable objects.

It has been argued that, since the dilaton field sits in one of the
hypermultiplets, the part of the Lagrangian involving vector
multiplets would not receive string loop correction, as the
consequence of the decoupling conjecture. Thus the BPS mass formula
is exact at the tree level string theory\cite{GK}. This corresponds to
use only the genus-zero partition function of the twisted
topological string theory as the prepotential for vector multiplets
in the effective supergravity theory.

However, more exact F-terms of the form $F_g W^{2g}$ are present in the
effective theory, all depend on powers of the $\mathcal{N}=2$
graviphoton superfield $W$, and explicitly depend on the high genus
partition function of the twisted topology string theory $F_g$~\cite{BCOV, AGNTI}. In the
effective action, they produce high derivative interactions, and
blackhole solutions and entropy countings have been studied
extensively to produce string loop corrected results~\cite{HDSUG, HDBH}.

We will notice that in the presence of constant graviphoton field
strength $\lambda^2$, these terms produce a series expansion in
$\lambda^2$ which corrects the prepotential for vector
supermultiplets. Combined with the corrected central charge formula,
we will see that they produce string loop corrected stability
conditions for BPS D-branes wrapping Calabi-Yau manifold. In
particular, this implies string loop corrected Hermitian-Yang-Mills
equation in the large volume limit for B-type branes, and loop
corrected special Lagrangian submanifold condition for A-type
branes. We will compare the known loop corrections to HYM equation
in heterotic string theory to support this claim.

It may seem strange that a string loop expansion should be
controlled by the graviphoton field strength. Similar phenomena has
occurred in the computation of the quantum corrected glueball
superpotential of the super-Yang-Mills theory, and has been studied
in~\cite{OV} and called C-deformation. The stability condition we
propose in this paper can be viewed as the effect on D-term in the
super-Yang-Mills theories obtained from the D-branes wrapping
Calabi-Yau manifold from such C-deformation, as in contrast to the
effect on superpotential in~\cite{OV}. The similarity between these
two is quire interesting.

The plan of the paper is as follows. After the introduction, we
examine the $D=4, \mathcal{N}=2$ effective supergravity action with
high derivative terms, and show that constant graviphoton field
strength deforms the moduli space geometry. Then after a conformal
field theory argument that graviphoton field strength can be
non-zero in string vacuum, we study a toy model of high derivative
pure supergravity, in which effective potential for graviphoton
field strength is present and constraints the value of graviphoton
field strength in the vacuum depending on the gravity background.
Then we study the central charge formula which includes the high
loop corrections. In the next section, we study the consequences of
the loop corrected D-brane stability conditions due to central charge
formula. For B-type branes in the large volume limit, loop
correction produces corrected Hermitian-Yang-Mills equation, which
matches the known loop corrected heterotic string result after type
I-heterotic duality. For A-type branes, the BPS brane is found to
wrap loop corrected special Lagrangian submanifolds. The last
section contains conclusions and discussions.

\section{Loop corrected central charge}

We first review the effective $D=4, \mathcal{N}=2$ supergravity action
 coupled with vector multiplets, together with the high derivative terms.
We will see the deformation of the moduli geometry in the presence of the
constant graviphoton field strength. We then look at the determinination of
the possible value of the background constant
graviphoton field strength in vacuum. It is this value controls the
genus expansion of the central charge, and the symplectic sections.
The expectation is that it can take arbitrary value, based on the
conformal field theory argument. However, in the effective action,
the supersymmetry requires the presence of potential for chiral
background, which fix it to certain value depending on the gravity
background and dynamics. Finally we present the deformed central charge formula.

\subsection{High derivative $D=4, \mathcal{N}=2$ effective theory and deformed moduli geometry}

We consider the $D=4$ effective theory of $\mathcal{N}=2$ conformal
supergravity obtained from type-II string theory compactification on
Calabi-Yau manifold. Upon gauge fixing, it produces the Poincare
supergravity. The fields in the theory include Weyl multiplet
$\mathcal{W}_{\alpha\beta}^{ij}$ whose lowest component is the
self-dual graviphoton field strength $T_{ab}^{ij}$, coupled with
(n+1) abelian vector multiplets $X^I (i=0, \cdots, n)$ whose lowest
components are scalars, which we again designated by $X^I$. The
hypermultiplets decouple from the vector multiplets, thus will not
concern us in the following discussion, except the fact that the
dilaton field sits in a hypermultiplet.

The effective action is constructed out of prepotential
\begin{align}
\int d^4x d^4 \theta \sum_g F_g(X^I) \mathcal{W}^{2g},
\end{align}
where the integral is over half of the $\mathcal{N}=2$ superspace
coordinates, and $F_g(X^I)$ is a holomorphic function of the vector
multiplets $X^I$'s. In the string theory compactification on
Calabi-Yau manifolds, $F_g(X^I)$ is equal to genus-g partition
function of the twisted topological string theory on Calabi-Yau
manifold~\cite{BCOV, AGNTI}.

The integral over half of the superspace coordinates will produce
various terms in the bosonic Lagrangian. We can absorb the $\theta$
coordinates by the Wely density $\mathcal{W}^2$, which gives
coupling terms of vector multiplets scalars with gravity, or we can
absorb the $\theta$ coordinates using vector multiples in
$F_g(X^I)$, which only the scalar part of the Weyl density
$\mathcal{W}^2$ appears. In the latter case, we obtain terms in the
action as a series expansion in the graviphoton field strength
$\lambda^2$
\begin{align}
\int d^4x d^4 \theta \sum_g F_g(X^I) \lambda^2, \qquad \lambda^2
\equiv \hat{A} = (T_{\mu\nu}^{ij}\epsilon_{ij})^2.
\end{align}
Notice that the action given by the $g=0$ term $F_0$ is the usual
prepotential for $\mathcal{N}=2$ field theory of vector multiplets, and it can
be also viewed as the result of setting $\lambda^2=0$ in the more
general action. Now it is obvious that in presence of the constant
graviphoton field strength, the prepotential is deformed to include
terms corresponding to high genus topological string amplitudes,
where the graviphoton field strength acts as deformation parameter
and controls the genus expansion.

The problem of constant graviphoton field strength in the vacuum
will be considered in the next section. Now we look at the
implications to the geometry of the moduli space.

The geometry of the target space $\mathcal{M}$ parameterized by the
scalars of the vector multiplets is a deformation of the usual $\mathcal{N}=2$ special
geometry. The full prepotential is again a homogeneous degree two
holomorphic function,
\begin{align}
F(cX^I, c^2\lambda^2) = c^2 F(X^I, \lambda^2),
\end{align}
where $X^I$ and $\lambda$ are degree-one sections of a line bundle
over the moduli space. $2(n+1)$ degree-one sections over the moduli
space can be obtained from the prepotential and again form a
section of a rank $2(n+1)$ symplectic vector bundle over the moduli
space $\mathcal{M}$,
\begin{align}
\begin{pmatrix} X^I \\ F_I(X^I, \lambda^2)  \end{pmatrix}.
\end{align}
Its symplectic inner product gives the deformed K\"{a}hler potential over
the moduli space, from which the deformed metric can be obtained by
taking appropriate derivatives
\begin{align}
e^{-\mathcal{K}} &= i[X^I \bar{F}_I(X^I, \lambda) - \bar{X}^I
F_I(X^I, \lambda)], \\
g_{i\bar{j}} &= \partial_i \bar{\partial}_{\bar{j}} \mathcal{K}.
\end{align}
The corresponding K\"{a}hler form on the moduli space is the deformation in
$H^2(\mathcal{M}, Z)$ of the one at $\lambda=0$.

Let us look at the deformation in terms of the moduli space of the
Calabi-Yau manifold. To be specific, we consider the moduli space of
the complex structures of the Calabi-Yau manifold $M$, which is
parameterized by the vector multiplets in the type IIA
compactifications. The moduli space of the Kahler structures can be
similarly obtained from the mirror symmetry.

At $\lambda=0$, the symplectic section is obtained as follows. Take
a symplectic basis of the homological three-cycles $\{A_\alpha,
B_\beta\}$, $\alpha, \beta=0,, 1, \cdots, n$ such that
\begin{align}
A_\alpha \cap B_\beta = \delta_{\alpha \beta}, \qquad A_\alpha \cap
A_\beta  =0, \qquad B_\alpha \cap B_\beta =0,
\end{align}
where we assume $H^3(M, Z)$ has real dimension $2(n+1)$. Take
holomorphic 3-form $\Omega$ on Calabi-Yau manifold, which is
uniquely defined up to a scale, the pairing between the homology and
the cohomology is given by the periods
\begin{align}
X^I =\int_{A_I} \Omega, \qquad F_I = \int_{B_I} \Omega,
\end{align}
which satisfy Picard-Fuchs equation in toric Calabi-Yau manifolds.

One can define a deformed 3-form $\Omega(\lambda)$ according to the
following property
\begin{align}
X^I =\int_{A_I} \Omega(\lambda), \qquad F_I(X^I, \lambda^2) =
\int_{B_I} \Omega(\lambda).
\end{align}
In general this prescription adds $H^{2,1}(M,C)$ piece to the holomorphic three-form.
We will argue that this three-form should be used to calibrate the loop corrected
special Lagrangian submanifolds.

Notice that here we assume the constant graviphoton field strength
without specifying its value. This works as long as we do not
consider the back reaction to the gravity. Certainly as will be evident in
the toy model of pure supergravity in the following section, its value should be
determined dynamically in specific gravity solutions.

\subsection{Non-zero constant graviphoton field strength}

To study the graviphoto photon field strength in vacuum, we first
look at the conformal field theory argument, basically
repeat the work in ~\cite{OV}.

Utilizing the covariant quantization of string theory, the
four-dimensional part of the worldsheet Lagrangian is
\begin{align}
\mathcal{L} = \frac{1}{2} \partial X^\mu\bar{\partial}X_\mu +
p_\alpha \bar{\partial} \theta^\alpha + p_{\dot{\alpha}}
\bar{\partial} \theta^{\dot{\alpha}} + \bar{p}_\alpha \partial
\bar{\theta}^\alpha + \bar{p}_{\dot{\alpha}} \partial
\bar{\theta}^{\dot{\alpha}},
\end{align}
where $p$'s are worldsheet (1,0) forms, $\bar{p}$'s are (0,1) forms,
and $\theta$ and $\bar{\theta}$'s are 0-forms. Turning on non-zero
self-dual graviphoton field strength $T_{\alpha \beta}$ adds to the
worldsheet the following term
\begin{align}
\int F^{\alpha\beta} p_\alpha \bar{p}_\beta.
\end{align}
The worldsheet theory is still conformal invariant, so it still
satisfies the string equation of motion.

From this argument, any value of the graviphoton field strength
would be a vacuum solution. However, effective potential for
graviphoton field strength could be generated from string theory
higher loop or non-perturbatively. Actually it will be present in
the effective theory, as we now turn to for a toy model of $D=4,
\mathcal{N}=2$ pure supergravity action.

\paragraph{Toy model: constant graviphoton field strength in pure supergravity}

Now we turn to a toy model of pure $D=4$ $\mathcal{N}=2$ supergravity
with high derivative corrections. One needs Weyl supermultiplet, one
vector supermultiplet and one hypermultiplet/nonlinear
multiplet/tensor multiplet (acts as compensator of the $SU(2)$ gauge
group in superconformal action) to match the degree of freedom of the Poincare supergravity
supermultiplet. The simplest action coupling to high
derivative terms is to assume
\begin{align}
F(X^0, \hat{A}) = \frac{-i}{4}(X^0)^2 +c\hat{A}.
\end{align}
Then the D-gauge fixing requires
\begin{align}
e^{-K} = |X^0|^2=1,
\end{align}
while the A-gauge fixing $X^0 =\bar{X}^0$ requires
\begin{align}
X^0=\pm 1.
\end{align}
We will choose $X^0=1$ in the following. So indeed the scalar field
in the vector multiplet is eliminated thorough the gauge fixing.

The bosonic part of the Lagrangian for this toy model, assuming the trivial behavior of
the fields such as $SU(2)$ gauge field, the hypermultiplet
field(decoupled from the vector multiplets), and the $U(1)$ gauge
field, after gauge fixing to obtain Poincare supergravity, is
\begin{align}
\begin{split}
8\pi e^{-1}\mathcal{L}_{\text{boson}} =& -\frac{1}{2} R +(32i c
{\mathcal{R}_{ab}^-}^{cd}
 {\mathcal{R}_{cd}^-}^{ab} + h.c. ) \\
 & + 8i(c-\bar{c})( \mathcal{D}_a T^{-ab} \mathcal{D}^c T^+_{cb}
 +\frac{1}{2}{R_a}^c T^{-ab}T^+_{cb}+\frac{1}{128} \hat{A}
\bar{\hat{A}}) \\
&+\frac{1}{8} (F^{-0} -\frac{1}{4}T^-)\cdot (F^{-0} -\frac{1}{4}T^-)
-\frac{1}{16} (F^{-0} -\frac{1}{4}T^-)\cdot T^- \\
& -\frac{i}{32} [\frac{-i}{4}(X^0)^2 +c\hat{A}] \bar{\hat{A}} +h.c.
\end{split}
\end{align}
Integrating out the vector field $F^{-0}$ identifies $T^-$ with the
graviphoton field strength
\begin{align}
F^{-0} =\frac{1}{2}T^{-0},
\end{align}
and also produces a potential linear potential term for $A$. The
Lagrangian reduces to
\begin{align}
\begin{split}
8\pi e^{-1}\mathcal{L}_{\text{boson}} =& -\frac{1}{2} R +(32i c
{\mathcal{R}_{ab}^-}^{cd}{\mathcal{R}_{cd}^-}^{ab} + h.c. )
+ 8i(c-\bar{c})\mathcal{D}_a T^{-ab} \mathcal{D}^c T^+_{cb}\\
& - \frac{1}{64} (A+\hat{A}) + 8i(c-\bar{c})( \frac{1}{2}{R_a}^c
T^{-ab}T^
+_{cb}+\frac{1}{128} \hat{A} \bar{\hat{A}})\\
\end{split}
\end{align}
Ignore the back reaction of the graviphoton field strength to the gravity, the
vacuum could be any conformal background including the flat space.
Now since there is a potential for the constant graviphoton field
strength, exactly of the form of $\phi^4$ potential for the scalar field,
while the scalar is the chiral scalar field constructed from the
graviphoton field strength $\hat{A}$. Notice that the Ricci-curvature
appears as part of the mass for graviphoton field strength, so fixing
graviphoton field strength is not decoupled from gravity. Even in the flat space background, the
field graviphoton field strength is nonzero, and fixed at
\begin{align}
\hat{A} = \frac{1}{4i(c-\bar{c})}.
\end{align}
In the dS or AdS space, the constant Ricci tensor acts as mass term
for graviphoton field strength, which fixes its value with dependence on
$c$ and the curvature of the spacetime. There is a critical value of
the curvature of the dS space above which the vacuum value of the
graviphoton field strength is zero.

\subsection{Deformed Central charge}

In Poincare supergravity without high derivative corrections, the
central charge for a blackhole carrying a set of electromagnetic
charges $(q_I, p^I)$ is defined as the integration of the spherical
part of the graviphoton field strength $T^{-\theta\phi}$ at spatial
infinity,
\begin{align}
Z = \frac{1}{4\pi}\int_{S^2_\infty} dS T^-_{\theta\phi}.
\end{align}
In this case, $T^-_{ab}$ is algebraically related to the symplectic
invariant $U(1)$ projection of the $N+1$ vector fields
\begin{align}
T^-_{ab} = F_I F^-_{ab} -X^IG^-_{ab}.
\end{align}
Thus there is no ambiguity to identify either as the the graviphoton
field strength. The central charge, after solving the equation of
motion for the vector fields and calculating the surface integral at
spatial infinity, turns out to be
\begin{align}
Z = e^{K/2} (p^IF_I -q_IX^I).
\end{align}

When high derivative terms are introduced through couplings to the
background supergravity fields, there is yet no strict derivation of
the central charge from the
supergravity action. A natural candidate is simply
\begin{align}
Z =e^{K(X^I, \lambda^2)/2} [p^IF_I(X^I, \lambda^2) -q_IX^I].
\end{align}
Indeed this is the only symplectic invariant $U(1)$ projected
expression, which reduces to the Poincare supergravity case when
$\lambda^2=0$. A strict derivation should proceed from analyzing the
supergravity action and obtained the supersymmetry algebra at
spatial infinity. This is quite difficult, since in high
derivative gravity the relation between $T^-_{ab}$ and the $U(1)$
projection of all the vector fields is not algebraic anymore. These two feidls
are related by complicated nonlinear equations which involve kinetic
terms for $T^{-ab}$ and back reaction to the metric is also present.
One possible point of view of to regard the $U(1)$ projection $F_I
F^-_{ab} -X^IG^-_{ab}$ as the physical graviphoton field strength,
while $T^-_{ab}$ as simply an auxiliary field whose value can only
be determined dynamically.

A large class of blackhole solutions preserving $D=4 \mathcal{N}=1$
symmetry are known, but asymptotically flat solutions seem to impose
certain restriction on the moduli parameters $X^I$'s at spatial
infinity, and non-asymptotically solutions are abundant. They
probably should still be called BPS blackholes as they are indeed
solutions to the equation of motion and preserve half of the bulk
supersymmetry. It is also suggested in~\cite{HMR} the
non-asymptotically flatness is due to the fact that $\alpha'$
corrections can not be ignored at spatial infinity and thus
$\lambda^2$ corrections should show up at spatial infinity. Note
that ADM mass for asymptotically AdS solutions is known, from which
the central charge can be obtained from the BPS condition
$|Z|=M$. This is important since that the class of BPS
blackhole solutions with asymptotically flatness seems to require
$\lambda^2=0$ at spatial infinity, while the $\mathcal{N}=2$
solution with $AdS_2\times S^2$ solutions when the moduli parameters
sit at attractor point has $\lambda^2 \neq 0$ at spatial infinity.
So asymptotic non-flat (maybe AdS or even dS) solutions should also
be included, where $\lambda^2$ dependent terms in (eqn. of central
charge) is expected to show up in the central charge formula. We
notice that at least the solutions with full $D=4$ $\mathcal{N}=2$
symmetries exhibit such property.

A microscopic definition from the moduli space parameters also
points to the correctness of above central charge formula. As we
have seen in the previous discussion, the moduli
space geometry in the presence of constant graviphoton field
strength, for A-type branes the loop corrections are summarized by a
$\lambda^2$ dependent three-form $\Omega(\lambda)$ from which symplectic sections are obtained . Thus the central
charge formula above is the obvious one.

Another confirmation of this expression as $\alpha'$ and $g_s$ loop
corrected central charge in the presence of constant graviphoton
field strength comes from its implications to D-brane stability
condition, which could be verified with statements from other
sources. This is what we will turn to in the next section.

\section{Deformed $\Pi$-stability condition and SLAG}

\subsection{Loop corrected $\Pi$-stability condition and HYM equation}

We first consider stability conditions for B-type BPS D-branes
wrapping even dimensional cycles of Calabi-Yau manifold. Because of
the decoupling of the vector multiplets and hypermultiplets, the
$\mathcal{N}=1$ supersymmetric configurations can be described in
two steps. In the first step, BPS B-type branes are identified as
objects in a category depends only on the complex structure of the
Calabi-Yau manifold. In the large volume limit, it is the derived
category of coherent sheaves on Calabi-Yau manifold. In the next step, each
brane configuration determines a central charge at each point in the K\"{a}hler moduli space via its topological
charges. Its phase defines a grading for each object in the category. Then $\Pi$-stability
criterion for a BPS brane configuration is defined by comparing the
phase of an object with that of all its sub-objects, which in the
large volume limit becomes the $\mu$ stability condition of the
holomorphic vector bundles.

This decoupling property is still preserved when $\alpha'$ and
string loop corrections are included, as dictated by the
$\mathcal{N}=2$ $D=4$ supersymmetry. Even in the presence of the
background graviphoton field strength, the vector multiplets and the
hypermultiplets are decoupled in the effective action, as long as
the vector fields are all abelian. Therefore we will single out the
second step in determining the stability of the branes
configuration, using the corrected central charge.

This in turn defines a loop corrected grading for type-B D-branes as
an object $E$ in the derived category of coherent sheaves\footnote{
The definition here differs from that in \cite{DFR} by a sign. It
results from different conventions of the signs in the periods. In
the large volume limit, it produces the same sign of the second term
in the grading as in ~\cite{DFR}. The differences in the grading is
in the sign of the constant which is trivial, and the third term
that we explicitly write out. This definition produces the right
correction as the grading from the improved charge expression.}
\begin{align}
\phi(E) =-\frac{1}{\pi} \text{arg} Z(E)
\end{align}
Type-B branes wrap $2k$ cycles. In the large volume limit, the
configuration can be described as a holomorphic vector bundle E,
whose Chern classes give the topological charges
\begin{align}
Q_6 =\text{rank}(E) =r; \qquad Q_4=\int_\Sigma c_1(E); \qquad
Q_2=\int_S ch_2(E); \qquad Q_0=\int_S ch_3(E),
\end{align}
which can be read from the following formula
\begin{align}
ch(E) = r+ c_1 +\frac{1}{2}(c_1^2 -2c_2) +\frac{1}{6}(c_1^3
-3c_1c_2+c_3) +\cdots.
\end{align}
They are associated with electric and magnetic charges separately,
\begin{align}
p^I = (Q_6, Q_4), \qquad q_I=(Q_2, Q_0).
\end{align}

As we have seen in the previous section, the periods paired with
various topological charges receive corrections as
$\lambda^2$-dependent terms. To compare with known results, we look
at the large volume limit where world sheet instanton contributions
are small and can be ignored. The full prepotential assumes the form
\begin{align}
F(t^i, \lambda) =\frac{1}{6} c_{ijk} t^it^jt^k -\frac{\lambda^2}{24}
c_{ai} t^i,
\end{align}
where $t^i$ parameterizes the complexified K\"{a}hler form
\begin{align}
B+iJ =\sum_i t^i \omega_i,
\end{align}
the $\omega_i$'s being a basis of K\"{a}hler two forms. The
coefficients $c_{ijk}$ are triple intersection forms,
\begin{align}
c_{ijk} = \int_M \omega_i\wedge \omega_j \wedge \omega_k.
\end{align}

The corrected prepotential produces the symplectic basis at large
volume limit ($J>>B$) as
\begin{align}
\begin{split}
\Pi_0 &=1; \\
\Pi_2^i &=t^i; \\
\Pi_4^i &=F_i(\lambda)= \frac{1}{2} c_{ijk} t^j t^k -\frac{\lambda^2}{24} c_{2i}, \\
\Pi_6 &=F(\lambda)=\frac{1}{6}c_{ijk}t^it^jt^k -\frac{\lambda^2}{24}
c_{2i}t^i.
\end{split}
\end{align}
This is based on the assumption that the the coordinate designation
should be reduced to the tree level result when set $\lambda=0$.
This should serve as a definition of the generalized special
geometry coordinates. Note the central charge is
\begin{align}
Z(E) = e^{K/2}(p^IF_I(X^I,\lambda) -q_I X^I),
\end{align}
where the factor $e^{K/2}$ is real and has not effect on the phase.
The expansion of the grading in the large volume limit is
\begin{align}
-\frac{1}{\pi} \text{Im} \log Z(E) = -\frac{3}{2} +\frac{6}{\pi}
\frac{\frac{1}{2}\int c_1(E)J^2 + \frac{\lambda^2}{24} \int
c_1(E)\wedge c_2(T) + \frac{1}{6}(c_1^3 -3c_1c_2+c_3)}{rV}+\cdots.
\end{align}
The $\alpha'$ and loop corrected slope $\mu$ is
\begin{align}
\mu(E) = \frac{1}{rV} [\int c_1(E)J^2 + \frac{\lambda^2}{12} \int
c_1(E)\wedge c_2(T) + \frac{1}{3}\int (c_1^3 -3c_1c_2+c_3)].
\end{align}
Now the $\alpha'$ and string loop corrected $\Pi$-stability
condition can be stated as follows: an object $E$ is $\Pi$-stable at
a point in K\"{a}hler moduli space if and only if for any subobject
$E'\subset E$ satsifies
\begin{align}
\phi(E') < \phi(E).
\end{align}

From this result, one notices the followings:
\begin{enumerate}
\item
It has been known for some time that the charge for D-branes in
curved space is
\begin{align}
Q(E) = ch(E) \sqrt{\hat{A}}
\end{align}
where $\hat{A}$ is the A-roof-genus of a spin manifold,
\begin{align}
\sqrt{\hat{A}} =1+ \frac{1}{12} c_2(M)+ \cdots
\end{align}
It is first discovered by the requirement of anomaly cancelation on
D-brane worldvolume. Combine this charge with the periods obtained
from tree level prepotential $F(X^I, \lambda^2=0)$ will reproduce the
grading formula exactly the same form as what we have derived above
in the large volume limit. This strongly suggests this anomaly
cancelation factor is from string loop. And our result can be seen
as a generalization of this term to the whole moduli space,
including all the worldsheet instanton corrections.

\item
From the Donaldson-Uhlenbeck-Yau theorem, the sufficient and
necessary condition for the existence of the Hermitian-Yang-Mills
equation
\begin{align}
F_{ab}=F_{\bar{a}\bar{b}}=0, \qquad g^{a\bar{b}}F_{a\bar{b}}=0
\end{align}
is that the slope of the vector bundle is zero
\begin{align}
\mu(E) =\frac{1}{rk(E)} \int_M J\wedge J \wedge c_1(E)=0,
\end{align}
which we will call DUY equation. Notice that this is the first term
in our formula. Now the $\alpha'$ and string loop corrected equation
in the large volume limit is
\begin{align}
\int_M c_1\wedge J\wedge J +\frac{\lambda^2}{12} \int_M c_1\wedge
c_2(M)+ \frac{1}{3}\int_M (c_1^3 -3c_1c_2+c_3)=0.
\end{align}
The stability condition for a D-brane configuration in type-I theory
should be the same, and after a duality map to heterotic string
theory can compare the result. The one-loop corrected stability
condition for an anomalous line bundle L in heterotic string theory
has been found out in~\cite{bhw} by considering gauge and mixed
anomaly cancelation, as
\begin{align}
\int_M c_1(L)\wedge J\wedge J -\frac{1}{2} l^2_s g_s^2 \int_M
(c_1(L)\wedge (\sum_k ch_2(V_k)+\sum_m a_m c_1^2(L_m) +\frac{1}{2}
c_2(M) )=0.
\end{align}
Using the anomaly cancelation condition for heterotic string
compactification
\begin{align}
\sum_k ch_2(V_k)+\sum_m a_m c_1^2(L_m) =-c_2(M)
\end{align}
this simplifies to
\begin{align}
\int_M c_1(L)\wedge J\wedge J +\frac{1}{4} l^2_s g_s^2 \int_M
c_1(L)\wedge  c_2(M)=0.
\end{align}
After the heterotic-type-I duality
\begin{align}
\begin{split}
e^{\phi_{10}^I} &= e^{-\phi_{10}^H}, \\
J^I &= J^H e^{-\phi_{10}^H},
\end{split}
\end{align}
this maps to an equation in type-I quantity
\begin{align}
\int_M c_1(L)\wedge J\wedge J +\frac{1}{4} l^2_s  \int_M
c_1(L)\wedge  c_2(M)=0,
\end{align}
which is exactly of the form of the first two terms in our equation.
Obviously, the MMMS equation, being a tree-level result, would only
include the first the the third term in our equation. The second
term, being a one-loop effect in heterotic string, is mapped to a
loop term in type-I. This picture matches perfectly to the fact that
all the higher derivative F-terms of the form $F_gW^{2g}$ which are
the origin for the correction to the stability conditions are
produced at one-loop level in heterotic string theory under the
$\mathcal{N}=2$ type II-heterotic duality~\cite{AGNTIII}. We regard
this as a strong evidence for our claim.
\end{enumerate}

Physically, the effect of the curved space in the large volume limit
can be interpreted as quantum two-brane charges of a six brane in
curved space produced by $c_2(M)$. This phenomena has appeared in
many places, and it is satisfactory that our claim in this limit
reproduces this result.

Finally, we point that what have obtained is not directly the loop
corrected Hermitian-Yang-Mills equation, but corrections to the
existence condition for its solution. If a generalized
Donaldson-Uhlenbeck-Yau theorem exists, the loop corrected condition
is presumed to give the sufficient and necessary condition for a
loop corrected HYM equation. It is in this sense that we claim to
have a correction to HYM equation. Certainly it would be interesting
to explicitly find out the correction terms to the
Hermitian-Yang-Mills equation.

\subsection{Loop corrected SLAG}

At tree-level, the mirror statement about D-brane stability on
A-branes concerns the phase of the restriction of the holomorphic
three form $\Omega$
\begin{align}
Re e^{i\theta} \Omega|_L =0,
\end{align}
for certain constant phase $\theta$. Here L is a submanifold of
middle dimension which satisfies
\begin{align}
\omega|_L =0.
\end{align}

It is obvious that after the inclusion of the string loop
correction, the SLAG condition should be corrected as
\begin{align}
\omega|_L =0, \qquad Re e^{i\theta} \Omega(\lambda)|_L =0,
\end{align}
using the corrected three-form $\Omega(\lambda)$ defined in the
previous section.

\section{Conclusions and discussions}

In this paper, we have argued that the high-derivative F-terms of
the form $F_gW^{2g}$, originated from high string loops, in the
presence of the constant graviphoton field strength would give a
loop corrected central charge for a configuration of D-branes. This
implies loop corrected stability condition for BPS D-branes, which
for B-type branes at large volume reduces to loop corrected
Hermitian-Yang-Mills equation, and for A-type branes reduces to a
deformed special Lagrangian submanifold condition. We have also show
that graviphoton field strength can be constant even in the presence
of effective potential as required by supersymmetry. We have
compared the loop corrected stability condition for B-type branes to
the various known results and obtained agreement.

There are several questions arise from this result. First, is the
string loop corrections to the D-stability condition present here complete? The
blackhole solutions and especially the blackhole entropy countings in
the presence of high derivative terms have been studied extensively.
The entropy countings have been checked to produce the right result in
the known examples. Although it is still a puzzle why this
particular class of terms should produce an exact result, the fact
that it is exact in known entropy counting lends support to the
conjecture that the D-brane stability condition as stated in this
paper is exact. Since physically these F-terms are produced by
integration out the effective 4d BPS particles obtained from
wrapping IIA 2-branes around 2-cycles in Calabi-Yau, one may wonder
the effects of integrating out branes wrapping higher dimensional
cycles. Certainly more understanding in this direction is desirable.

Second, the similar stability problem can be studied in the dual
picture of the D-branes wrapping Calabi-Yau and spanning the spacetime.
It is known to show up as D-term in the super-Yang-Mills theory on
the D-branes. A microscopy string theory calculation of the D-term in the
presence of constant graviphoton field strength should be able to
verify our claim.

Finally, a probable lesson from the result of this paper is that
supersymmetry breakings on the D-brane could be tied up with
supergravity background, and could be quite complicated. The
implications to supersymmetry breaking and moduli fixing problem in
string phenomenology are certainly worth further study.

This work is supported by NSF grant PHY-0244811 and DOE grant
DE-FG01-00ER45832.



\end{document}